2024

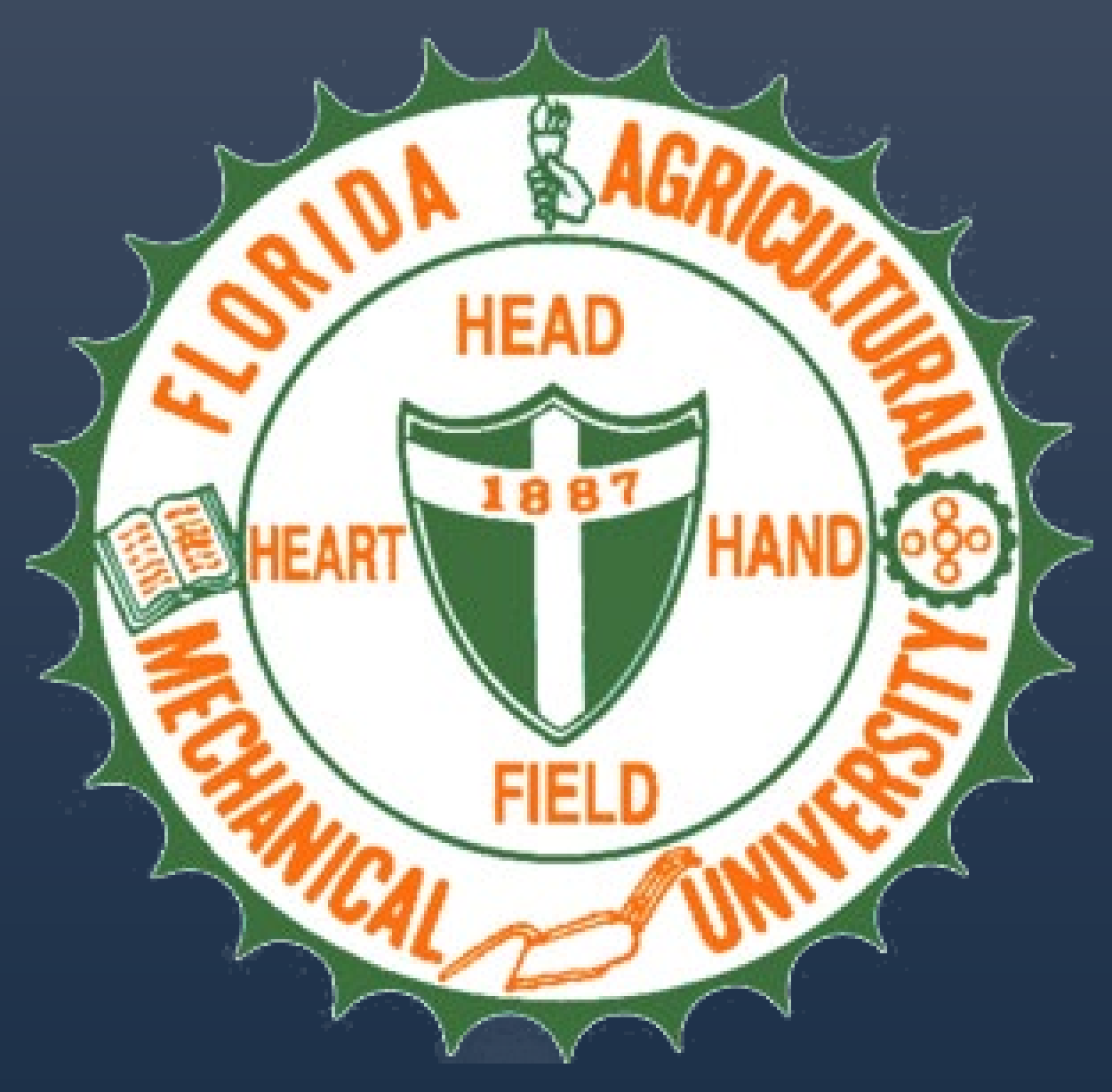

# RURAL COMMUTE PATTERNS

PREPARED BY:

MOHAMED KHALAFALLA, DOREEN KOBELO, THOBIAS SANDO, JANEROZA MATYENYI

# EXECUTIVE SUMMARY

Transportation provides access to employment opportunities and essential services such as healthcare services, while urban areas have various transportation options, the situation differs in rural areas. Rural residents often have longer commute distances, limited access to public transit, and extended waiting times for public transportation if they exist, which can significantly impact their access to vital services and job opportunities. This study used data from the 2017 NHTS survey to examine the commuting patterns in rural areas by utilizing multinomial logistic regression to determine how various factors impact the choice of mode of transport in rural areas. Findings from this study revealed a higher dependency, 92.1%, on using personal vehicles when making trips in rural areas. Multinomial logistic regression results showed that socio-demographics, household, and trip characteristics affect the mode of transport used in a trip. Older adults, females, and individuals with higher education levels than high school graduates are less likely to use public transit when making trips. For household characteristics, the availability of vehicles in a household and households with higher income levels have lower probabilities of making a trip using public transit. Longer trip distances reduce the likelihood of a trip using active commuting modes such as walking and biking. These findings provide insights into understanding the transportation behaviors in rural areas and provide knowledge to be used in the planning and developing of transportation projects to promote equitable and accessible transportation in rural areas.

**Table of Contents**

## List of Figures

## List of Tables

# 1. INTRODUCTION

Transportation plays a vital role in the growth and development of rural communities. Having reliable and efficient transportation options allows residents to access employment, education, healthcare services, and connections to larger metropolitan areas. However, rural areas have limited transportation options, characterized by infrequent or no public transit services and long travel distances (US Department of Transportation, 2019). This lack of transportation options hinders economic opportunities and leads to limited access to services such as health services that impact people's health and isolate communities (Syed et al., 2013).

Travel patterns in rural communities across the United States are not as well understood as those in urban and suburban areas. This gap in knowledge on rural commuting patterns and their variations across different types of rural areas poses significant challenges for planning and developing transportation systems that meet the mobility needs of rural residents. which negatively impacts their mobility, accessibility, and overall quality of life.

This study analyzes the commute patterns of people in rural areas and how different factors, including socio-demographic, trip, and household characteristics, impact the choice of mode of commute in rural areas. It also takes into account how commute characteristics differ across census regions.

# 2. LITERATURE REVIEW

## 2.1. Rural and Urban Areas

The National Household Travel Survey (NHTS) utilizes the classification of urban and rural areas as defined by the U.S. Census Bureau. This distinction relies on population thresholds, land use, population density, and housing characteristics. Rural areas encompass territories, housing, and populations that do not fall within urban or urban cluster designations. On the other hand, urban areas are characterized by densely developed regions consisting of residential, commercial, and other non-residential urban land uses (Ratcliffe et al., 2016).

According to the U.S. Census Bureau criteria, an urban area constitutes a densely populated core of census blocks with a minimum housing unit density of 2000, a population density of at least 1000 individuals per square mile, and a total population of at least 5000 people (Census Bureau, 2016). The demographic landscape of the U.S. reflects a significant urbanization trend, with approximately 80% of the population residing in urban areas, a substantial increase from 64% in 1950. Despite this demographic shift, urban land occupies only 3% of the total land area, while rural areas encompass 97% of the landmass. Surprisingly, despite their vast expanse, rural areas accommodate roughly 20% of the country's population (American Community Survey, 2015). Table 1 presents the populations and percentage of people living in rural areas in the United States, Puerto Rico, and Island Areas.

***Table 1. 2020 Census Urban Areas by the Numbers (Census.gov, 2022)***

| | United States | Puerto Rico | Island Areas |
|---|---|---|---|
| Total number of 2020 Census Urban Areas | 2,613 | 26 | 7 |
| Total urban population | 265,149,027 | 3,018,908 | 294,319 |
| Percent population living within urban areas | 80.0% | 91.9% | 87.1% |
| Total rural population | 66,300,254 | 266,966 | 43,702 |
| Percent population living within rural areas | 20.0% | 8.1% | 12.9% |

## 2.2. Census Regions and Divisions

The United States has a large geographical area with states and counties having diverse physical and cultural geography. The Census Bureau groups the states and counties into various regions and divisions representing different sections of the United States. The current configuration of Census regions was established in 1910, following modifications to the initial areas introduced in 1850.

The Census Bureau classifies states and counties into four regions: Northeast, West, Midwest, and South. These regions are further classified into nine divisions: Pacific, Mountain, West South Central, East South Central, South Atlantic, West North Central, East North Central, Middle Atlantic, and New England, as illustrated in **Figure 1**.

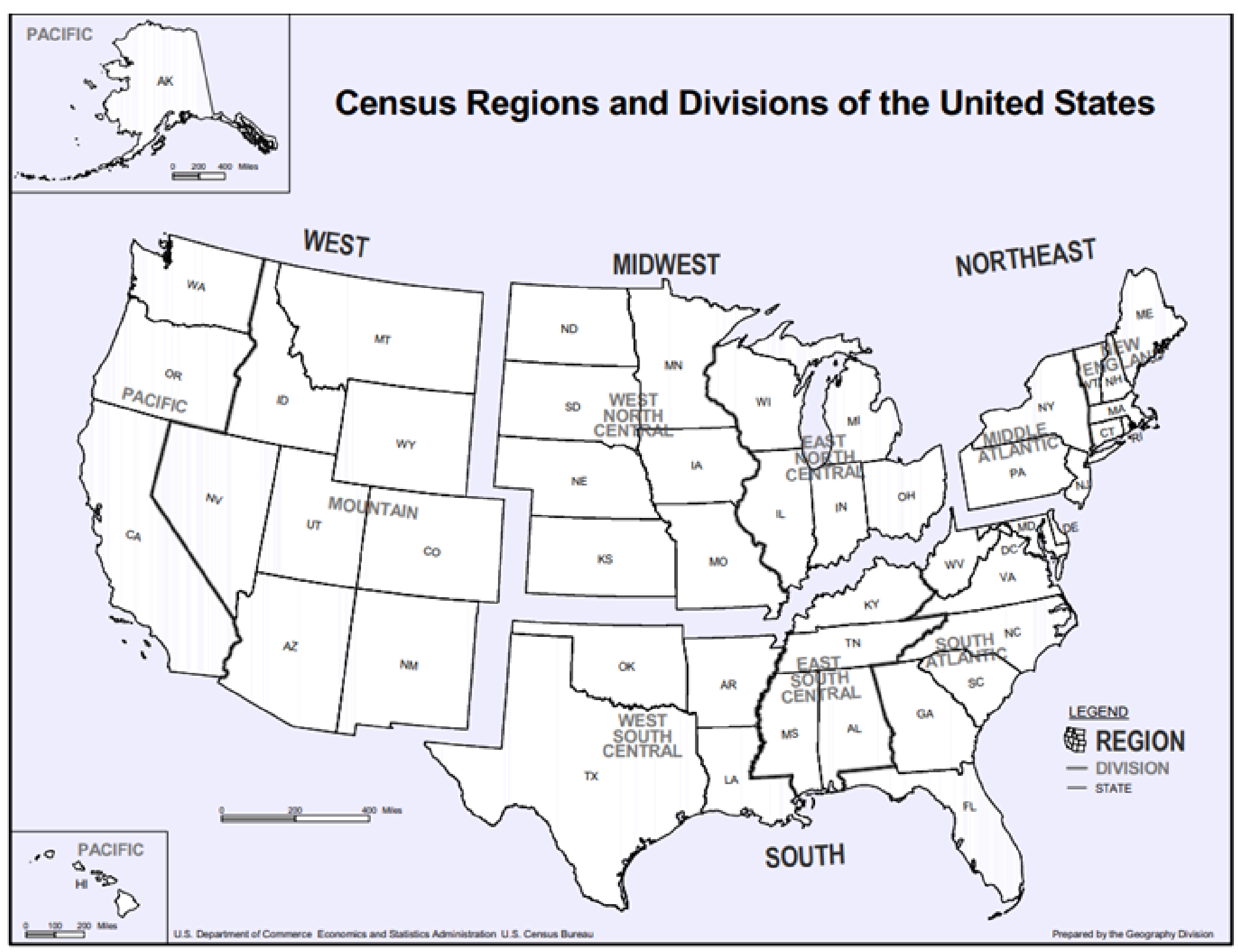

*Figure 1. Census Region and Divisions of the United States*

### 2.3. Data Sources and Description

The National Household Travel Survey is conducted by the Federal Highway Administration every 5 to 7 years, and most recently, it was conducted in 2017. The NHTS data is the source of data on travel behavior and trends at the national level. The survey collects data on demographic characteristics such as age, gender, and race, data on travel-related characteristics including vehicle types, daily trip making, and distance traveled within the trips, and data on socioeconomic factors such as the income of households. The 2017 survey was conducted between April 2016 and April 2017. It includes information on 923,572 trips made by 264,234 individuals aged five or older in 129,696 US households (Federal Highway Administration, 2018).

The data consists of four files: the household file, which contains data collected once for the household; the person file, which contains data collected once for each interviewed household member; the vehicle file, which has data related to household vehicles; and the trip file, which has data items collected for each trip made. This study used trip file data to analyze the travel behavior of rural residents. The trip file includes all trips made on the assigned date by persons of age five and older; the 24-hour travel day started at 4:00 a.m. of the assigned day and ended at 3:59 a.m. of the following day. Figure 2 illustrates an example of trips made on the trip day that are recorded in the trip file (Federal Highway Administration, 2018).

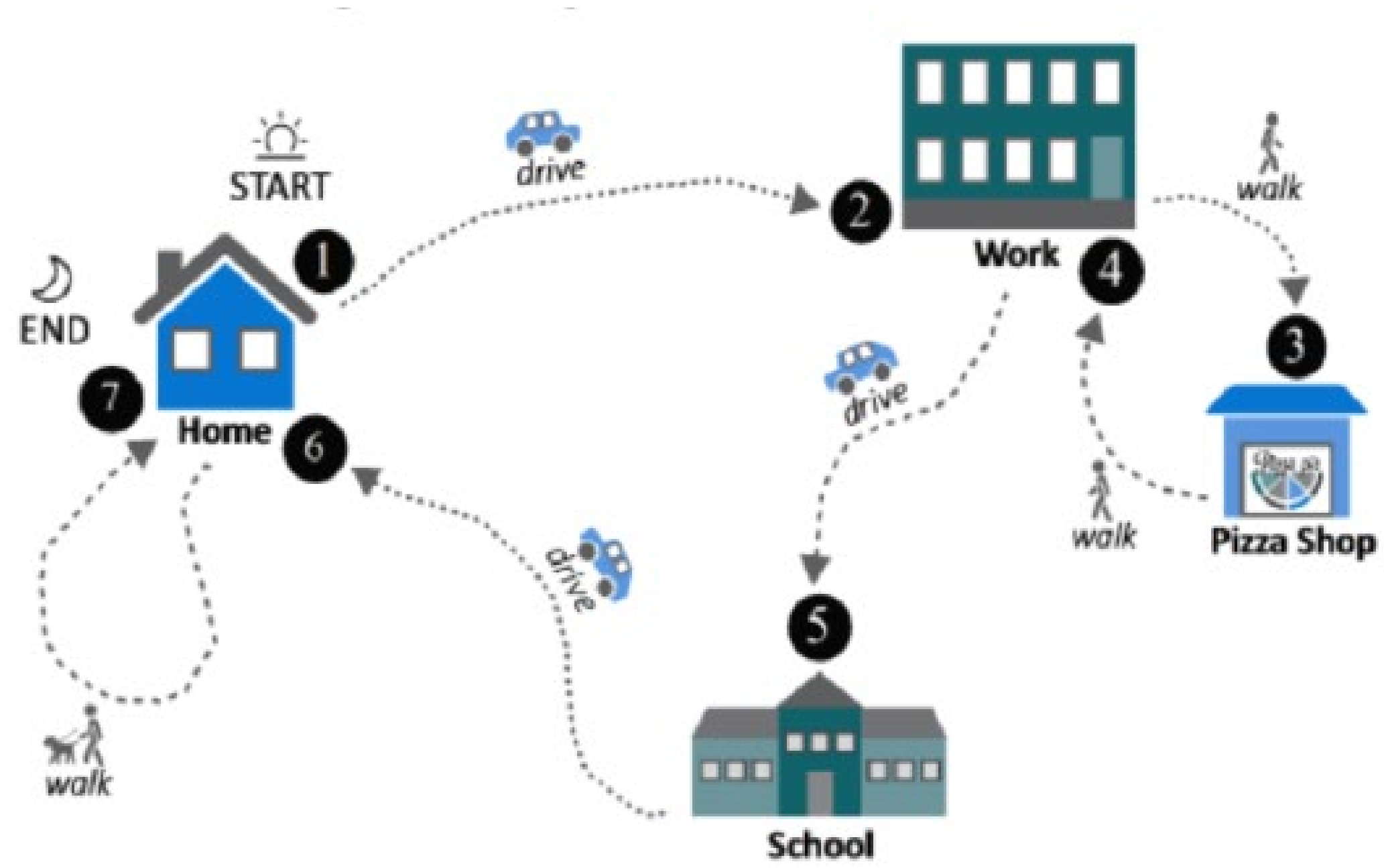

*Figure 2. Illustrations of trips made on a trip day*

### 2.4. Multinomial Logistic Regression

Multinomial Logistic Regression is a statistical regression model that is used to analyze the association between categorical outcome variables with more than two categories and a set of independent variables (Cheng Hua et al., 2021). In most cases, multinomial regression is used to study how people or organizations make choices, the model assumes that variation in the characteristics of decision-makers determines variation in choice outcomes (Bernasco & Block, 2012). In this study, multinomial logistic regression was used to predict how different characteristics affect the choice of the mode of transport used in a trip.

To apply the MNL model, the reference category should be defined and used as the baseline or the comparison group (Fan et al., 2016). In this study, the reference category is the personal vehicle. The results were interpreted by examining each variable's odds ratio and p-values. An odds ratio is mainly used to analyze the effect of independent variables on the response variable. It is calculated by the equation (1).

$$odds\ ratio = exp^{coefficient} \qquad (1)$$

# 3. RESULTS AND DISCUSSION

This section provides an overview of the study's findings and discussions. It focuses on analyzing variable correlations and presenting and discussing results from the multinomial logistic regression model.

### 3.1. Descriptive Statistics

*Mode of transport*

Figure 3 presents the mode of transport used in a trip in rural areas; findings show a strong dependence on personal vehicles when making a trip, with 92.1% of all trips made by personal vehicles. Active commuting modes, walking, and biking, account for 5.9% and 0.3% of trips, respectively, while other modes comprise 1.2%. Public transport usage is notably low, representing only 0.4% of all trips.

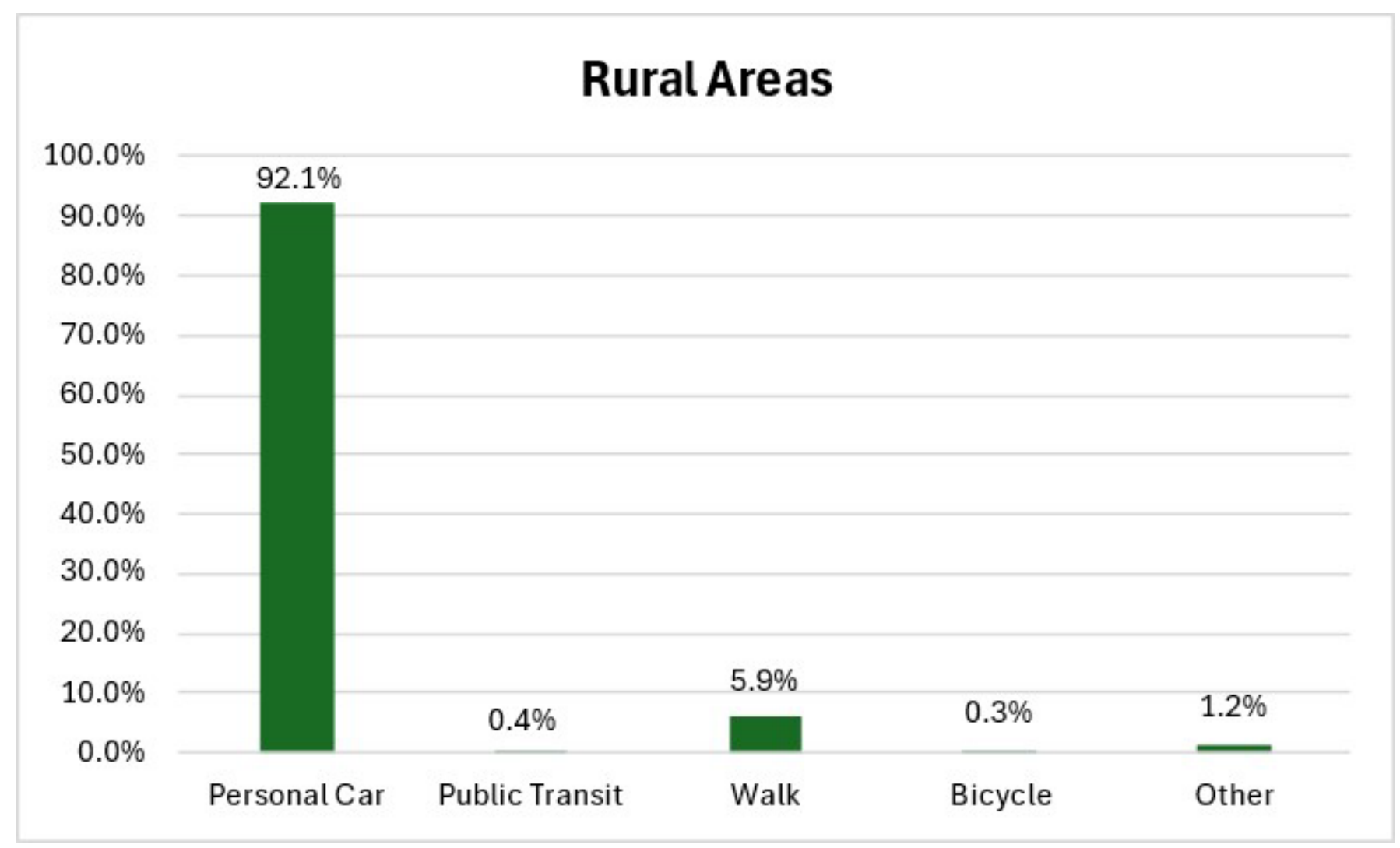

*Figure 3. Mode of transport used in a trip*

*Household Size*

The household size distribution in rural areas, as presented in Figure 4, reveals that two-person households constitute 48% of the total households. Households with three and four members consist of 14.7% and 14.3% of the households, respectively. Larger households with five or more members represent 9.8%, while single-person households account for 13.3%. of the total. This indicates that rural areas are likely characterized by small families, an aging population living alone, and young couples.

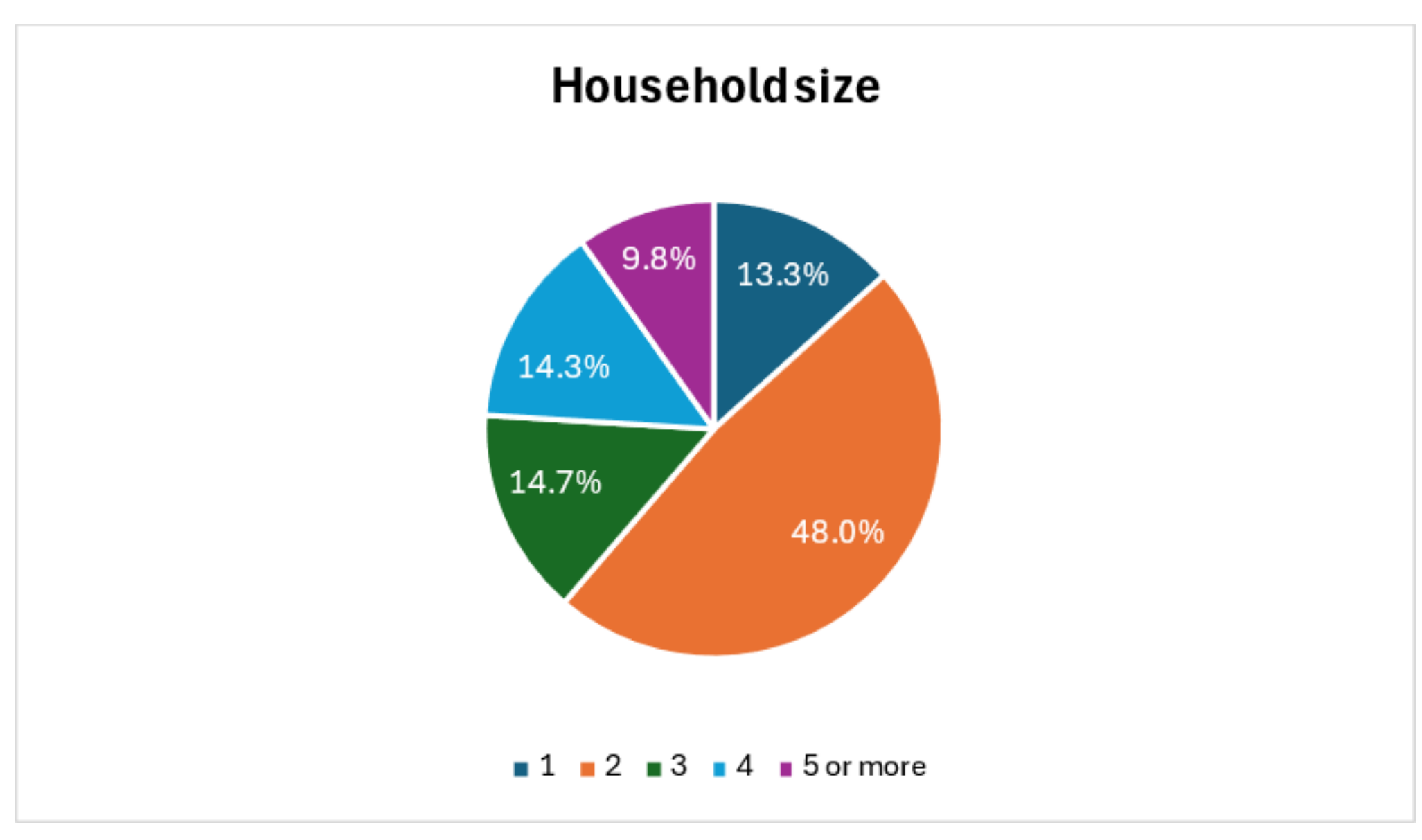

*Figure 4.Household size distribution*

*Household Vehicles*

Vehicle ownership patterns in rural households, presented in Figure 5, show the dependency on personal vehicles for transportation in rural areas. The majority of the households 39.3% have two vehicles, 25.5% possess three vehicles, and 19.9% possess four or more vehicles. Only 14.5% of households have a single vehicle, while 0.8% have no vehicle at all. This high rate of vehicle ownership in households (84.7% owning two or more vehicles) shows the perceived necessity of car ownership in rural areas.

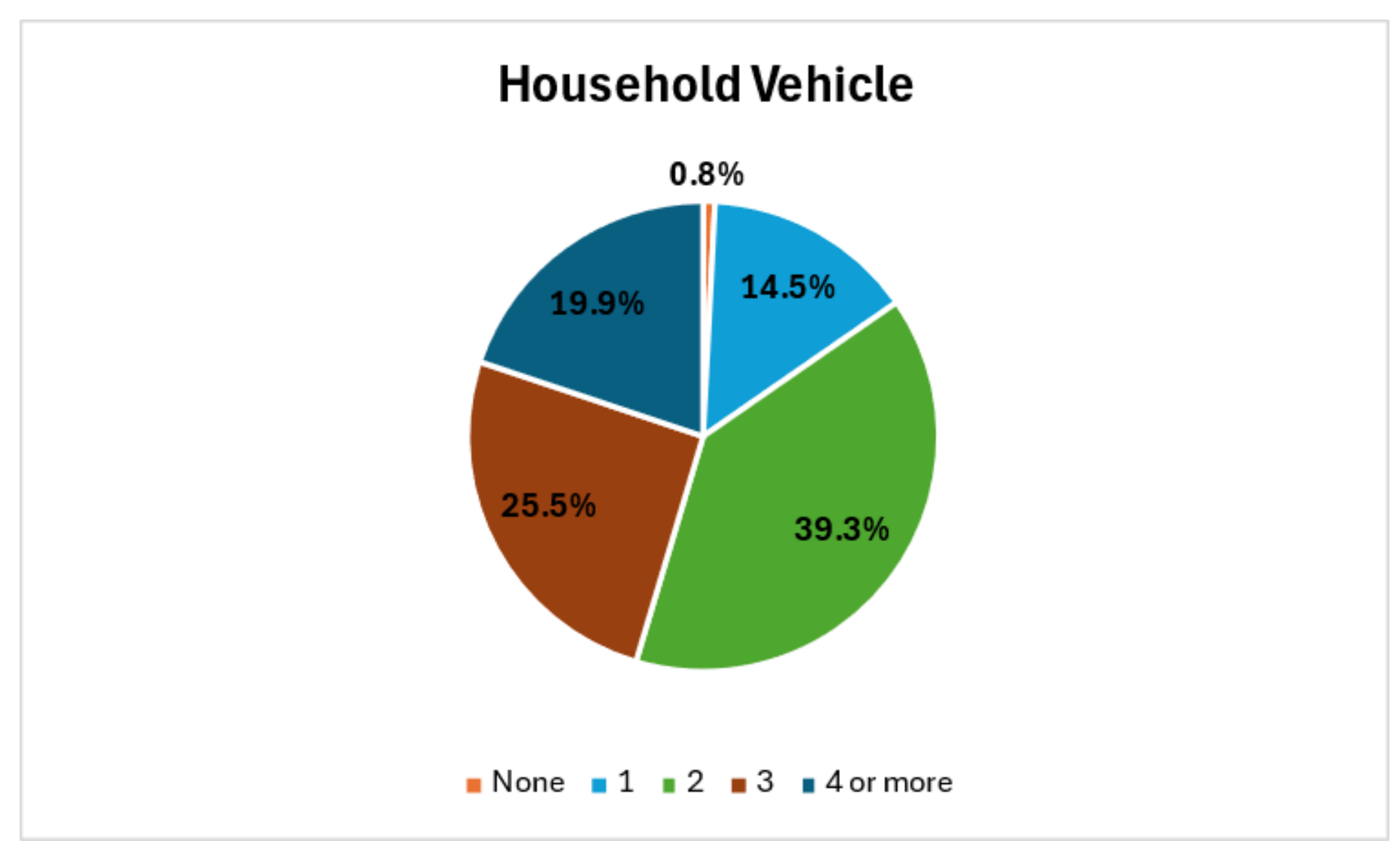

*Figure 5. Percentage of vehicles present in a household*

*Trip Destination*

Figure 6 presents the trip destinations originating from rural areas; the majority, 69%, of trips are within rural areas. Small towns are the second most common destination, accounting for 19.2% of trips, while 5.9% of the trips are to second cities and 5.3% to suburban areas. Urban areas are the least common destination, with 0.6%. This highlights the interconnectedness of rural areas with nearby towns. centers,

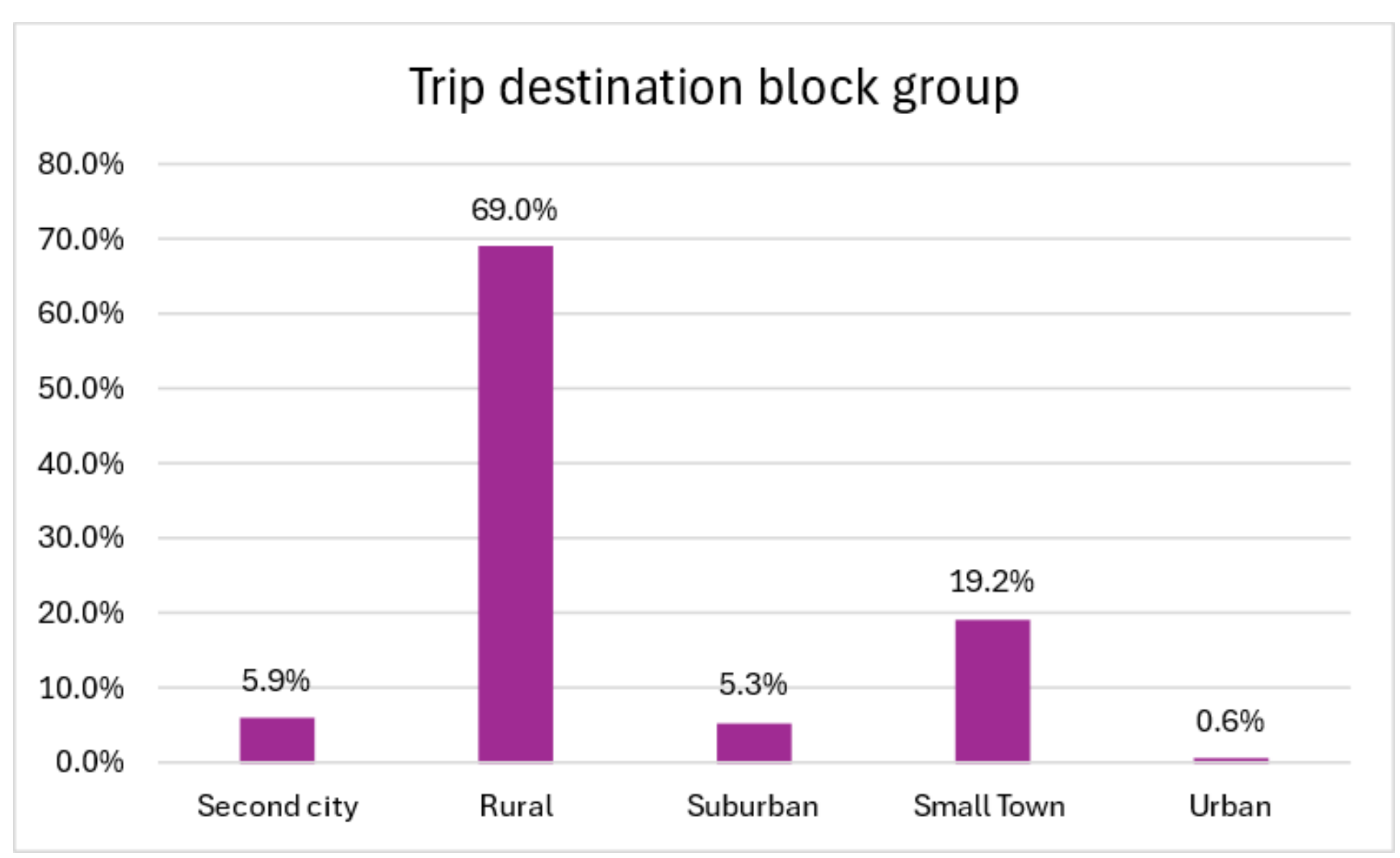

*Figure 6. Trip destination block group*

**3.2. Multinomial Logistic Regression Results**

Results from the Multinomial logistic regression are presented in Table 2. The modes of transport were categorized into public transit, personal vehicle, walking, bicycle, and other. The category other consists of modes of transport that could not be classified into the above groups, such as golf carts, motorcycles, rideshare, and those referred to as others in the data. Modes of transport, such as airplanes and ferries, were excluded from the study's analysis.

The interpretation of results was based on the odds ratio, and the p-value was used to determine the statistical significance of the variable. A variable was found to be statistically significant at the 95% CI if it had a p-value of less or equal to 0.05. The fitted model had a prediction accuracy of 91.68%. The mode of transport personal vehicle was set as the base category for the dependent variables; results are discussed concerning this category.

***Table 2. Multinomial logistic regression results***

| Variable | Transit | | Walk | | Bicycle | | Other | |
|---|---|---|---|---|---|---|---|---|
| | **Odds Ratio** | **Pvalue** | **Odds Ratio** | **Pvalue** | **Odds Ratio** | **Pvalue** | **Odds Ratio** | **Pvalue** |
| ***Age*** | | | | | | | | |
| 11-30 | | | | | | | | |
| 31-50 | **0.76** | **0.06** | 0.57 | P < 0.01 | 0.47 | P < 0.01 | 1.99 | P < 0.01 |
| 51-70 | 0.49 | P < 0.01 | 0.54 | P < 0.01 | 0.34 | P < 0.01 | 1.94 | P < 0.01 |
| 71 and above | 0.42 | P < 0.01 | 0.31 | P < 0.01 | 0.05 | P < 0.01 | **1.15** | **0.29** |
| ***Gender*** | | | | | | | | |
| Male | | | | | | | | |
| Female | 0.71 | P < 0.01 | 0.88 | P < 0.01 | 0.3 | P < 0.01 | 0.23 | P < 0.01 |
| ***Education level*** | | | | | | | | |
| Highschool graduate or less | | | | | | | | |
| College degree | 0.77 | 0.04 | 1.23 | P < 0.01 | **0.75** | **0.06** | 0.78 | P < 0.01 |
| Bachelor's degree | 1.13 | 0.32 | 1.87 | P < 0.01 | 1.55 | P < 0.01 | 0.53 | P < 0.01 |
| ***Race*** | | | | | | | | |
| White | | | | | | | | |
| Black | 4.33 | P < 0.01 | **0.92** | **0.35** | **1.53** | **0.11** | 0.68 | 0.03 |
| Hispanic | 1.96 | P < 0.01 | 0.76 | 0.01 | 1.63 | 0.05 | **0.84** | **0.33** |
| Other | **1.1** | **0.71** | **1.16** | **0.06** | **1.07** | **0.8** | **1.05** | **0.74** |
| ***Household Income*** | | | | | | | | |
| less than 25,000 | | | | | | | | |
| 25000-49999 | 0.45 | P < 0.01 | **0.97** | **0.57** | 0.7 | 0.04 | 1.3 | 0.01 |
| 50,000-74999 | 0.54 | P < 0.01 | 1.12 | 0.04 | 0.52 | P < 0.01 | 1.27 | 0.02 |
| 75,000-99999 | 0.35 | P < 0.01 | 1.21 | P < 0.01 | 0.56 | P < 0.01 | 1.31 | 0.02 |
| 100,000 and more | **1.14** | **0.39** | 1.7 | P < 0.01 | **0.94** | **0.72** | 1.48 | P < 0.01 |
| ***Household Size*** | 0.87 | P < 0.01 | 0.88 | P < 0.01 | 0.75 | P < 0.01 | 0.92 | P < 0.01 |
| ***Household Vehicle*** | **0.65** | P < 0.01 | 0.94 | P < 0.01 | **0.92** | **0.06** | 1.14 | P < 0.01 |
| ***Trip purpose*** | | | | | | | | |
| Work | | | | | | | | |
| Non-home based | 0.67 | P < 0.01 | 0.43 | P < 0.01 | 0.44 | P < 0.01 | 1.75 | P < 0.01 |
| Shopping | 0.26 | P < 0.01 | 0.62 | P < 0.01 | 0.61 | 0.05 | 0.74 | P < 0.01 |
| Recreation | 0.6 | P < 0.01 | 4.69 | P < 0.01 | 5.99 | P < 0.01 | 2 | P < 0.01 |
| Other | 0.57 | P < 0.01 | 3.32 | P < 0.01 | 2.03 | P < 0.01 | 0.89 | 0.27 |
| ***Trip Day*** | | | | | | | | |
| Weekend | | | | | | | | |
| Weekday | 1.58 | P < 0.01 | **1.04** | **0.24** | 1.36 | 0.02 | 1.21 | 0.01 |
| ***Trip miles*** | 1.00 | P < 0.01 | 0.18 | P < 0.01 | 0.64 | P < 0.01 | 1 | P < 0.01 |
| ***Census Region*** | | | | | | | | |
| Northeast | | | | | | | | |
| Midwest | 0.17 | P < 0.01 | 0.64 | P < 0.01 | 1.94 | P < 0.01 | 1.22 | 0.02 |
| South | 0.23 | P < 0.01 | 0.55 | P < 0.01 | 0.71 | 0.02 | **0.98** | **0.74** |
| West | 0.38 | P < 0.01 | **0.92** | **0.06** | **1.14** | **0.44** | **0.94** | **0.54** |

#### 3.2.1. Personal Characteristics

*Age*

The results show a significant correlation between age and mode of transport choice for rural trips. Older adults show a decreased likelihood of using public transit compared to younger individuals aged 11-30, with those aged 31-50 being 0.76 times less likely, 51-70 being 0.49 times less likely, and individuals 71 and above 0.42 times less likely to use public transit than the youngest group. This trend suggests a growing preference for personal vehicles among older rural residents, potentially due to increased car ownership or reduced comfort with public transportation options.

A similar age-related decline is observed in active transportation modes such as walking and biking. Compared to the 11-30 age group, individuals aged 31-50 are 0.57 times less likely to walk and 0.47 times less likely to bike. These findings indicate that as rural residents age, they increasingly favor personal vehicles overactive transportation modes. This shift may be attributed to various factors, including health-related limitations that make walking or biking more challenging for older adults, as well as the convenience and comfort offered by personal vehicles in rural settings

*Gender*

Gender significantly impacts the mode of transport chosen for rural trips, as evidenced by the p-values being less than 0.05 for several categories (Table 2). Compared to males, females are 0.71 times less likely to use public transit, 0.88 times less likely to walk, 0.30 times less likely to bike, and 0.23 times less likely to use other modes of transportation besides personal vehicles, public transit, walking, and biking.

The lower likelihood of females using active transportation modes such as walking and biking could be due to safety concerns, lack of infrastructure catering to their needs, or sociocultural factors influencing travel behavior. Additionally, the reduced use of other modes by females may stem from limited access to services like ride-sharing or specialized transportation options tailored for women in rural areas. These gender disparities highlight the need for improved safety measures, better infrastructure, and more inclusive transportation services to accommodate the needs of female travelers in rural communities.

*Education Level*

Education level plays a significant role in the mode of transport chosen for rural trips, as indicated by the p-values in Table 2. Compared to individuals with a high school education or less, those with a college degree are 0.77 times less likely to use public transit and 0.78 times less likely to use other modes of transport. However, college graduates are 1.23 times more likely to walk.

Individuals with a bachelor's degree or higher education are 1.87 times more likely to walk and 1.55 times more likely to bike compared to those with a high school education or less. Additionally, they are 0.53 times less likely to use other modes of transport.

These findings suggest that higher educational attainment is associated with a greater likelihood of engaging in active transportation modes such as walking and biking. This trend could be attributed to better health awareness, greater environmental consciousness, or access to resources that promote active commuting. Conversely, the reduced reliance on public transit and other modes of transport among higher-educated individuals may be linked to their greater financial capacity to own and maintain personal vehicles.

*Race*

Race significantly influences the mode of transport chosen for rural trips, as shown by the p-values in Table 2. Compared to Whites, Black Americans are 4.33 times more likely to use public transit, 0.68 times less likely to use other modes of transport, and do not show a significant difference in walking. Hispanics are 1.96 times more likely to use public transit, 0.76 times less likely to walk, and 1.63 times more likely to bike compared to Whites.

These results highlight racial disparities in transportation mode choices. The higher likelihood of using public transit among Black Americans and Hispanics may indicate greater reliance on public transportation due to socioeconomic factors such as lower vehicle ownership rates and income levels. The lower likelihood of walking among Hispanics could be due to cultural preferences or safety concerns. The increased likelihood of biking among Hispanics suggests that biking might be a more accessible and affordable mode of transportation for this group.

#### 3.2.2. Household Characteristics

*Household Size*

The size of a household plays a crucial role in determining the mode of transport for rural trips, as evidenced by the p-values in Table 2. With each additional household member, the likelihood of using public transit decreases by 0.87 times, walking by 0.88 times, biking by 0.75 times, and other modes of transport by 0.92 times.

This trend indicates that larger households are less inclined to use public transit and active transportation modes. The reasons for this could include logistical challenges in coordinating travel for multiple members and a higher reliance on personal vehicles to meet the diverse needs of a larger family. Financial constraints may also limit the ability of larger households to afford multiple transportation options, making personal vehicles a more practical choice for managing daily travel needs.

*Household Vehicles*

The number of vehicles in a household significantly impacts the mode of transport chosen for rural trips, as shown by the p-values in Table 2. An increase in the number of household vehicles decreases the likelihood of using public transit and walking by 0.65 times and 0.94 times, respectively. Conversely, the presence of more household vehicles increases the likelihood of using other modes of transport by 1.14 times.

This suggests that households with more vehicles are more likely to rely on personal and other motorized modes of transport, reducing their dependence on public transit and active transportation. The availability of household vehicles offers greater flexibility and convenience, allowing residents to cover longer distances and access a wider range of destinations more efficiently.

### 3.2.3. Trip Characteristics

*Trip Purpose*

The characteristics of a trip, such as its purpose and timing, significantly influence the choice of transport mode. For work-related trips, public transit is more likely to be used. In contrast, non-home-based, shopping, recreation, and other purpose trips are 0.67, 0.26, 0.60, and 0.57 times less likely to be made using public transit, respectively.

Recreational trips, on the other hand, show a higher likelihood of involving active transportation modes. These trips are 4.69 times more likely to involve walking and 5.99 times more likely to involve biking, indicating the importance of having safe and accessible infrastructure for walking and biking to support recreational activities.

*Trip Day*

The day of the week also influences the mode of transport chosen. Weekday trips are 1.58 times more likely to involve public transit compared to weekend trips. Additionally, weekday trips are 1.36 times more likely to involve biking and 1.21 times more likely to involve other modes of transport.

This suggests that public transit and active transportation modes are more commonly used for weekday activities, likely due to the regular commuting needs for work and school. Understanding these patterns can help in optimizing public transit schedules and improving infrastructure to support active transportation during the weekdays.

*Trip Length*

Trip length significantly impacts the mode of transport chosen for rural trips. As the length of the trip increases, the likelihood of using public transit and other modes increases, while the likelihood of using active modes such as walking and biking decreases. Specifically, for each additional mile, a trip is 1.01 times more likely to use public transit and other modes, 0.64 times less likely to involve biking and 0.18 times less likely to involve walking.

This indicates that longer trips are better suited for public transit or motorized modes of transport, whereas shorter trips are more feasible for active transportation modes. These findings are similar to those of Yang et al. 2018, suggesting that the trip distance and cost increase the chances of using public transit when making a trip. This underscores the need for integrated transportation solutions that offer efficient public transit for longer trips while promoting active transportation for shorter distances.

### 3.2.4. Census Regions

Regional differences also play a significant role in transportation mode choice. Individuals in the Midwest, South, and West regions are 0.17, 0.23, and 0.38 times less likely to use public transit in their trips than those in the Northeast.

Additionally, trips in the Midwest are 1.22 times more likely to be made using other modes of transport, 1.94 times more likely to involve biking, and 0.64 times less likely to involve walking. In the South, trips are 0.55 times less likely to involve walking and 0.71 times less likely to involve biking.

These regional variations highlight the importance of considering local context when developing transportation policies and infrastructure. Tailored strategies that address each region's unique needs and characteristics can help promote sustainable and efficient transportation options across different parts of the country.

### 3.3. Differences Across Census Regions in Rural Areas

A separate multinomial model was created for each of the four census regions. The results from the multinomial logistic regression models for the four census regions—Northeast, Midwest, South, and West—reveal distinct patterns and variations in transportation mode choices across the United States. This section compares and contrasts these regional differences to comprehensively understand how various factors influence transportation preferences in each region.

The comparison of results across the four regions highlights significant regional variations in transportation mode choice influenced by age, gender, education, race, household income, household size, trip purpose, trip day, and trip length. The Northeast shows a higher reliance on public transit and active transportation modes, likely due to better transit infrastructure and urban density. The Midwest and South regions have lower public transit usage and higher reliance on personal vehicles, reflecting these areas' more rural and spread-out nature. The West shows unique patterns, particularly with significant racial disparities in mode choice, indicating potential cultural and infrastructural differences. Table 3 summarizes the findings of the models for each region. The discussion is based on 95% confidence level results. The odds ratios associated with the summary presented in Table 3 are shown in Appendices A1 through A4.

*Table 3. Summary of findings of factors affecting mode choice in rural areas for each census region*

| Age | |
|---|---|
| Northeast | Older adults are less likely to use public transit, walk, or bike compared to younger individuals, with significant decreases in likelihood for each successive age group. |
| Midwest | Similar to the Northeast, older adults are less likely to walk or bike, but the use of public transit does not significantly decrease with age. |
| South | Older adults show a significant decrease in walking and biking, with the 71 and above age group showing the most substantial reduction. |
| West | Older adults are less likely to use public transit, walk, or bike, with a dramatic decrease in biking for those aged 71 and above. |
| **Gender** | |
| Northeast | Females are less likely than males to use public transit, walk, or use other modes of transport. |
| Midwest | Females are less likely to bike or use other modes, with no significant difference in walking or public transit use compared to males. |
| South | Females are less likely to walk, bike, or use other modes of transport, with public transit usage not significantly different. |
| West | Females are less likely to bike or use other modes, with no significant difference in walking or public transit use compared to males. |
| **Education level** | |
| Northeast | Higher educational attainment is associated with a higher likelihood of walking and using other modes of transport. |
| Midwest | College graduates are less likely to use public transit or bike, while individuals with a bachelor's degree are more likely to walk. |
| South | Higher educational attainment increases the likelihood of walking and using other modes of transport, while public transit usage decreases slightly with higher education. |
| West | Individuals with higher education are more likely to walk, with no significant difference in biking or public transit use. |
| **Race** | |
| Northeast | Black Americans and Hispanics are significantly more likely to use public transit compared to Whites, with Hispanics also more likely to use other modes of transport. |
| Midwest | Black Americans are significantly more likely to walk, while Hispanics show a decreased likelihood of using public transit. |
| South | Black Americans and Hispanics are more likely to use public transit and bike, with Black Americans less likely to walk. |
| West | Black Americans are significantly more likely to walk and bike, while other racial groups show varying preferences for walking and other modes. |
| **Household Income** | |
| Northeast | Higher household income decreases the likelihood of using public transit but increases the likelihood of walking and using other modes. |
| Midwest | Higher income is associated with less public transit use and more biking, while walking shows mixed results. |
| South | Higher income increases the likelihood of walking and using other modes, with higher income households less likely to use public transit. |
| West | Higher income decreases the likelihood of using public transit and other modes, with mixed results for walking and biking. |
| **Household Size** | |
| Northeast | Larger households are less likely to use public transit, walk, or bike. |
| Midwest | Larger households show a decreased likelihood of walking and using other modes, with no significant difference in biking. |
| South | Larger households are less likely to walk, bike, or use other modes. |
| West | Larger households are less likely to use public transit, walk, or bike, with no significant difference in using other modes. |
| **Trip Purpose** | |
| Northeast | Work trips are more likely to involve public transit, while recreational trips are more likely to involve walking and biking. |
| Midwest | Work trips favor public transit, while non-home-based trips are more likely to use other modes. |
| South | Work trips are more likely to use public transit, while recreational trips show a higher likelihood of walking and biking. |
| West | Recreational trips are more likely to involve walking, with shopping trips less likely to use public transit. |
| **Trip Day** | |
| Northeast | Weekday trips are more likely to use public transit compared to weekends. |
| Midwest | Weekday trips are less likely to use public transit. |
| South | Weekday trips show a significantly higher likelihood of using public transit, biking, and other modes. |
| West | Weekday trips are more likely to use other modes, with public transit usage decreasing on weekends. |
| **Trip Length** | |
| Northeast | Longer trips are more likely to use public transit and other modes, with a decrease in walking and biking. |
| Midwest | Longer trips are more likely to use public transit and other modes, with a decrease in walking and biking. |
| South | Longer trips increase the likelihood of using public transit and other modes, while decreasing walking and biking. |
| **West** | Longer trips show a similar trend of increasing public transit and other mode usage, with decreased walking and biking. |

# 4. CONCLUSIONS

This study used the 2017 NHTS data to examine the travel behavior of people in rural areas and employed multinomial logistic regression to analyze how the mode choice of transportation is affected by socio-demographics, household, and trip characteristics. Results from the MNL show that older adults (51-70 and 71+ age groups) are 0.49 and 0.42 times less likely to use public transit compared to younger individuals. Females are 0.71 times less likely to use public transit and 0.30 times less likely to bike. College degree holders are 1.23 times more likely to walk, and individuals with a bachelor's degree or higher are 1.87 times more likely to walk and 1.55 times more likely to bike. Black Americans and Hispanics are 4.33 and 1.96 times more likely to use public transit than Whites.

Work-related trips are more likely to involve public transit, while recreational trips are 4.69 times more likely to involve walking and 5.99 times more likely to involve biking. Weekday trips are 1.58 times more likely to use public transit compared to weekend trips. Regional differences show that individuals in the Northeast are significantly more likely to use public transit than those in the Midwest, South, and West. As trip length increases, the likelihood of using public transit rises, while walking and biking become less feasible, suggesting the integration of multimodal options for longer trips.

These findings emphasize the need for targeted strategies, including developing age-friendly transportation options, promoting active commuting through educational programs and infrastructure improvements, and ensuring equitable transit services for minority communities. By considering these factors, transportation systems can become more inclusive, efficient, and responsive to the needs of diverse populations, enhancing mobility for all rural communities.

# APPENDICES

## Appendix A1

### Multinomial Logistic Regression for Northeast Region

| Variable | Transit | | Walk | | Bicycle | | Other | |
|---|---|---|---|---|---|---|---|---|
| | Odds Ratio | Pvalue | Odds Ratio | Pvalue | Odds Ratio | Pvalue | Odds Ratio | Pvalue |
| **Age** | | | | | | | | |
| 11-30 | | | | | | | | |
| 31-50 | **0.67** | **0.08** | 0.62 | P < 0.01 | **1.03** | **0.95** | 2.12 | P < 0.01 |
| 51-70 | **0.69** | **0.09** | 0.57 | P < 0.01 | **0.96** | **0.92** | 1.73 | 0.01 |
| 71 and above | 0.56 | 0.04 | 0.29 | P < 0.01 | 0.22 | 0.01 | **1.48** | **0.13** |
| **Gender** | | | | | | | | |
| Male | | | | | | | | |
| Female | 0.72 | 0.01 | **0.94** | **0.24** | **0.33** | **P < 0.01** | 0.42 | P < 0.01 |
| **Education level** | | | | | | | | |
| Highschool graduate or less | | | | | | | | |
| College degree | **0.96** | **0.85** | 1.28 | P < 0.01 | **1.54** | **0.20** | 0.56 | P < 0.01 |
| Bachelor's degree or more | **1.11** | **0.56** | 1.81 | P < 0.01 | 2.19 | 0.01 | 0.47 | P < 0.01 |
| **Race** | | | | | | | | |
| White | | | | | | | | |
| Black | 5.01 | P < 0.01 | **2.38** | **0.08** | 0.00 | P < 0.01 | 0.00 | P < 0.01 |
| Hispanic | 3.71 | P < 0.01 | **0.84** | **0.52** | **1.80** | **0.43** | 2.89 | P < 0.01 |
| Other | **0.82** | **0.64** | **0.95** | **0.77** | **1.08** | **0.89** | 1.81 | 0.03 |
| **Household Income** | | | | | | | | |
| less than 25,000 | | | | | | | | |
| 25000-49999 | 0.31 | P < 0.01 | 0.76 | 0.01 | **0.63** | **0.17** | **1.36** | **0.14** |
| 50,000-74999 | 0.56 | 0.01 | 0.82 | 0.05 | 0.40 | 0.02 | **1.35** | **0.15** |
| 75,000-99999 | 0.44 | P < 0.01 | 0.80 | 0.05 | 0.29 | 0.01 | **0.78** | **0.32** |
| 100,000 and more | **1.04** | **0.85** | 1.38 | P < 0.01 | 0.46 | 0.04 | **1.22** | **0.37** |
| **Household Size** | **1.05** | **0.46** | 0.90 | P < 0.01 | **1.02** | **0.86** | **1.04** | **0.52** |
| **Household Vehicle** | 0.58 | P < 0.01 | 0.87 | P < 0.01 | 0.51 | P < 0.01 | **1.04** | **0.38** |
| **Trip purpose** | | | | | | | | |
| Work | | | | | | | | |
| Non-home based | 0.55 | P < 0.01 | 0.63 | P < 0.01 | **0.62** | **0.28** | 2.53 | P < 0.01 |
| Shopping | 0.31 | P < 0.01 | **0.75** | **0.06** | **0.67** | **0.40** | **1.12** | **0.64** |
| Recreation | 0.54 | 0.01 | 4.90 | P < 0.01 | 2.91 | 0.01 | 2.01 | P < 0.01 |
| Other | 0.53 | P < 0.01 | 3.76 | P < 0.01 | **2.07** | **0.09** | **1.26** | **0.34** |
| **Trip Day** | | | | | | | | |
| Weekend | | | | | | | | |
| Weekday | 1.73 | P < 0.01 | **1.12** | **0.06** | **1.12** | **0.65** | **1.00** | **0.99** |
| **Trip miles** | 1.00 | P < 0.01 | 0.25 | P < 0.01 | 0.92 | P < 0.01 | 1.00 | P < 0.01 |

## Appendix A2

## Multinomial Logistic Regression for Midwest Region

| Variable | Transit | | Walk | | Bicycle | | Other | |
|---|---|---|---|---|---|---|---|---|
| | Odds Ratio | Pvalue | Odds Ratio | Pvalue | Odds Ratio | Pvalue | Odds Ratio | Pvalue |
| **Age** | | | | | | | | |
| 11_30 | | | | | | | | |
| 31-50 | **1.37** | **0.57** | 0.76 | 0.04 | 0.51 | 0.02 | 1.54 | 0.04 |
| 51-70 | **1.24** | **0.70** | 0.72 | 0.01 | 0.49 | 0.01 | **1.39** | **0.10** |
| 71 and above | **1.24** | **0.73** | 0.42 | P < 0.01 | 0.02 | P < 0.01 | 0.22 | P < 0.01 |
| **Gender** | | | | | | | | |
| Male | | | | | | | | |
| Female | **1.63** | **0.07** | **0.98** | **0.76** | 0.46 | P < 0.01 | 0.24 | P < 0.01 |
| **Education level** | | | | | | | | |
| Highschool graduate or less | | | | | | | | |
| College degree | 0.41 | 0.03 | **1.11** | **0.27** | 0.39 | P < 0.01 | 0.73 | 0.01 |
| Bachelor’s degree or more | **1.36** | **0.35** | 1.63 | P < 0.01 | **1.19** | **0.43** | 0.42 | P < 0.01 |
| **Race** | | | | | | | | |
| White | | | | | | | | |
| Black | 0.02 | P < 0.01 | **3.08** | **0.07** | 0.01 | P < 0.01 | **3.10** | **0.07** |
| Hispanic | 0.00 | P < 0.01 | **0.54** | **0.27** | 0.01 | P < 0.01 | 0.01 | P < 0.01 |
| Other | **2.45** | **0.22** | 0.50 | 0.03 | **1.22** | **0.78** | **0.35** | **0.12** |
| **Household Income** | | | | | | | | |
| less than 25,000 | | | | | | | | |
| 25000-49999 | 0.35 | 0.04 | **1.13** | **0.35** | **0.96** | **0.93** | **1.33** | **0.23** |
| 50,000-74999 | **0.39** | **0.07** | **1.19** | **0.19** | **1.67** | **0.28** | **0.98** | **0.94** |
| 75,000-99999 | **0.86** | **0.75** | 1.51 | P < 0.01 | 2.79 | 0.03 | **1.55** | **0.07** |
| 100,000 and more | **1.25** | **0.64** | 1.74 | P < 0.01 | 2.65 | 0.03 | **1.44** | **0.12** |
| **Household Size** | **0.79** | **0.13** | 0.91 | 0.01 | **0.86** | **0.11** | 0.81 | P < 0.01 |
| **Household Vehicle** | 0.77 | 0.05 | **0.98** | **0.49** | **0.89** | **0.15** | 1.18 | P < 0.01 |
| **Trip purpose** | | | | | | | | |
| Work | | | | | | | | |
| Non-home based | 0.51 | 0.04 | 0.55 | P < 0.01 | 0.18 | P < 0.01 | 2.64 | P < 0.01 |
| Shopping | 0.13 | P < 0.01 | 0.64 | 0.01 | 0.25 | P < 0.01 | **0.78** | **0.27** |
| Recreation | 0.34 | 0.03 | 5.03 | P < 0.01 | 1.76 | 0.02 | **1.39** | **0.15** |
| Other | **0.47** | **0.07** | 3.12 | P < 0.01 | 0.41 | P < 0.01 | **1.15** | **0.50** |
| **Trip Day** | | | | | | | | |
| Weekend | | | | | | | | |
| Weekday | 0.42 | P < 0.01 | **0.99** | **0.90** | **1.24** | **0.38** | **1.13** | **0.38** |
| **Trip miles** | 1.01 | P < 0.01 | 0.26 | P < 0.01 | 0.82 | P < 0.01 | 1.01 | P < 0.01 |
| **Census Division** | | | | | | | | |
| D4 | 0.21 | 0.01 | 0.68 | P < 0.01 | **0.98** | **0.94** | **1.03** | **0.82** |

## Appendix A3

### Multinomial Logistic Regression for West Region

| Variable | Transit | | Walk | | Bicycle | | Other | |
|---|---|---|---|---|---|---|---|---|
| | Odds Ratio | Pvalue | Odds Ratio | Pvalue | Odds Ratio | Pvalue | Odds Ratio | Pvalue |
| **Age** | | | | | | | | |
| 11_30 | | | | | | | | |
| 31-50 | **0.56** | **0.14** | **1.05** | **0.74** | **1.02** | **0.97** | **0.85** | **0.56** |
| 51-70 | 0.35 | P < 0.01 | **0.77** | **0.06** | **0.82** | **0.65** | **1.12** | **0.67** |
| 71 and above | 0.20 | P < 0.01 | 0.56 | P < 0.01 | 0.01 | P < 0.01 | 0.73 | 0.32 |
| **Gender** | | | | | | | | |
| Male | | | | | | | | |
| Female | **0.99** | **0.95** | **0.97** | **0.68** | 0.22 | P < 0.01 | 0.26 | P < 0.01 |
| **Education level** | | | | | | | | |
| Highschool graduate or less | | | | | | | | |
| College degree | **1.69** | **0.14** | **0.95** | **0.63** | **1.23** | **0.54** | **1.11** | **0.56** |
| Bachelor's degree or more | **1.75** | **0.13** | 1.75 | P < 0.01 | **1.13** | **0.74** | **0.83** | **0.34** |
| **Race** | | | | | | | | |
| White | | | | | | | | |
| Black | 0.00 | P < 0.01 | 7.04 | 0.03 | 13.40 | 0.04 | 0.00 | P < 0.01 |
| Hispanic | **1.56** | **0.27** | **1.28** | **0.10** | **1.01** | **0.99** | **0.66** | **0.23** |
| Other | **0.49** | **0.25** | 1.40 | 0.01 | **1.31** | **0.56** | 0.41 | 0.03 |
| **Household Income** | | | | | | | | |
| less than 25,000 | | | | | | | | |
| 25000-49999 | 0.01 | 0.01 | **0.91** | **0.39** | **1.06** | **0.90** | 0.64 | 0.06 |
| 50,000-74999 | 0.45 | 0.05 | **1.04** | **0.75** | **0.58** | **0.31** | 0.44 | P < 0.01 |
| 75,000-99999 | **0.96** | **0.92** | **1.08** | **0.55** | **1.60** | **0.32** | **0.79** | **0.35** |
| 100,000 and more | **0.69** | **0.29** | **1.08** | **0.50** | **1.91** | **0.13** | **0.73** | **0.16** |
| **Household Size** | 0.75 | 0.02 | 0.80 | P < 0.01 | 0.65 | P < 0.01 | **1.05** | **0.44** |
| **Household Vehicle** | **0.89** | **0.22** | 1.08 | P < 0.01 | **1.07** | **0.44** | 1.11 | 0.03 |
| **Trip purpose** | | | | | | | | |
| Work | | | | | | | | |
| Non-home based | **0.97** | **0.92** | 0.44 | P < 0.01 | 0.00 | P < 0.01 | **1.31** | **0.23** |
| Shopping | 0.20 | P < 0.01 | 0.56 | P < 0.01 | 0.00 | P < 0.01 | **0.79** | **0.38** |
| Recreation | 0.31 | 0.05 | 4.84 | P < 0.01 | 0.00 | P < 0.01 | **1.68** | **0.06** |
| Other | **0.86** | **0.71** | 2.54 | P < 0.01 | 0.00 | P < 0.01 | 0.50 | 0.03 |
| **Trip Day** | | | | | | | | |
| Weekend | | | | | | | | |
| Weekday | 0.56 | 0.02 | **1.12** | **0.14** | **1.02** | **0.95** | 1.73 | P < 0.01 |
| **Trip miles** | 1.00 | P < 0.01 | 0.26 | P < 0.01 | 0.93 | P < 0.01 | **1.00** | **0.49** |
| Census Division | | | | | | | | |
| Division 9 | **0.84** | **0.52** | **1.11** | **0.21** | **0.66** | **0.10** | **0.96** | **0.82** |

## Appendix A4

### Multinomial Logistic Regression for South Region

| Variable | Transit | | Walk | | Bicycle | | Other | |
|---|---|---|---|---|---|---|---|---|
| | Odds Ratio | Pvalue | Odds Ratio | Pvalue | Odds Ratio | Pvalue | Odds Ratio | Pvalue |
| **Age** | | | | | | | | |
| 11-30 | | | | | | | | |
| 31-50 | **0.66** | **0.12** | 0.56 | P < 0.01 | 0.27 | P < 0.01 | 1.78 | P < 0.01 |
| 51-70 | **0.63** | **0.07** | 0.59 | P < 0.01 | 0.11 | P < 0.01 | 1.69 | P < 0.01 |
| 71 and above | **0.66** | **0.16** | 0.39 | P < 0.01 | 0.02 | P < 0.01 | 1.32 | 0.12 |
| **Gender** | | | | | | | | |
| Male | | | | | | | | |
| Female | **0.77** | **0.10** | 0.91 | 0.05 | 0.01 | 0.03 | 0.29 | P < 0.01 |
| **Education level** | | | | | | | | |
| Highschool graduate or less | | | | | | | | |
| College degree | **0.93** | **0.71** | 1.33 | P < 0.01 | 0.41 | 0.02 | **0.92** | **0.36** |
| Bachelor's degree or more | **0.75** | **0.19** | 1.72 | P < 0.01 | **1.15** | **0.67** | 0.61 | P < 0.01 |
| **Race** | | | | | | | | |
| White | | | | | | | | |
| Black | 3.39 | P < 0.01 | 0.84 | 0.07 | 2.24 | 0.01 | **0.80** | **0.16** |
| Hispanic | **1.17** | **0.71** | **0.93** | **0.58** | 4.45 | P < 0.01 | **0.72** | **0.19** |
| Other | **0.75** | **0.57** | 1.71 | P < 0.01 | **0.50** | **0.40** | 1.41 | 0.06 |
| **Household Income** | | | | | | | | |
| less than 25,000 | | | | | | | | |
| 25000-49999 | **0.88** | **0.57** | **0.86** | **0.06** | 0.00 | 0.05 | 1.48 | 0.01 |
| 50,000-74999 | 0.47 | 0.04 | **0.92** | **0.32** | 0.20 | P < 0.01 | 1.60 | P < 0.01 |
| 75,000-99999 | **0.68** | **0.34** | 1.32 | P < 0.01 | 0.00 | P < 0.01 | 1.68 | P < 0.01 |
| 100,000 and more | 3.59 | P < 0.01 | 1.50 | P < 0.01 | **1.40** | **0.34** | 1.57 | P < 0.01 |
| **Household Size** | **1.04** | **0.60** | 0.95 | 0.04 | 0.64 | P < 0.01 | 0.84 | P < 0.01 |
| **Household Vehicle** | 0.41 | P < 0.01 | 0.92 | P < 0.01 | 0.77 | 0.02 | 1.18 | P < 0.01 |
| **Trip purpose** | | | | | | | | |
| Work | | | | | | | | |
| Non-home based | 1.77 | 0.03 | 0.28 | P < 0.01 | 0.02 | P < 0.01 | 1.45 | P < 0.01 |
| Shopping | **0.70** | **0.28** | 0.30 | P < 0.01 | 0.00 | P < 0.01 | 0.61 | P < 0.01 |
| Recreation | **1.32** | **0.46** | 3.44 | P < 0.01 | 0.00 | P < 0.01 | 2.56 | P < 0.01 |
| Other | **0.89** | **0.73** | 2.39 | P < 0.01 | 0.00 | P < 0.01 | **0.91** | **0.54** |
| **Trip Day** | | | | | | | | |
| Weekend | | | | | | | | |
| Weekday | 32.58 | P < 0.01 | 1.21 | P < 0.01 | 3.39 | P < 0.01 | 1.56 | P < 0.01 |
| **Trip miles** | 1.00 | P < 0.01 | 0.25 | P < 0.01 | 0.86 | P < 0.01 | 1.01 | P < 0.01 |
| **Census Division** | | | | | | | | |
| Division 6 | **0.97** | **0.94** | **0.85** | **0.25** | **0.86** | **0.80** | **1.82** | P < 0.01 |

# Appendix A5

## Visualized Odds Ratios

Gender

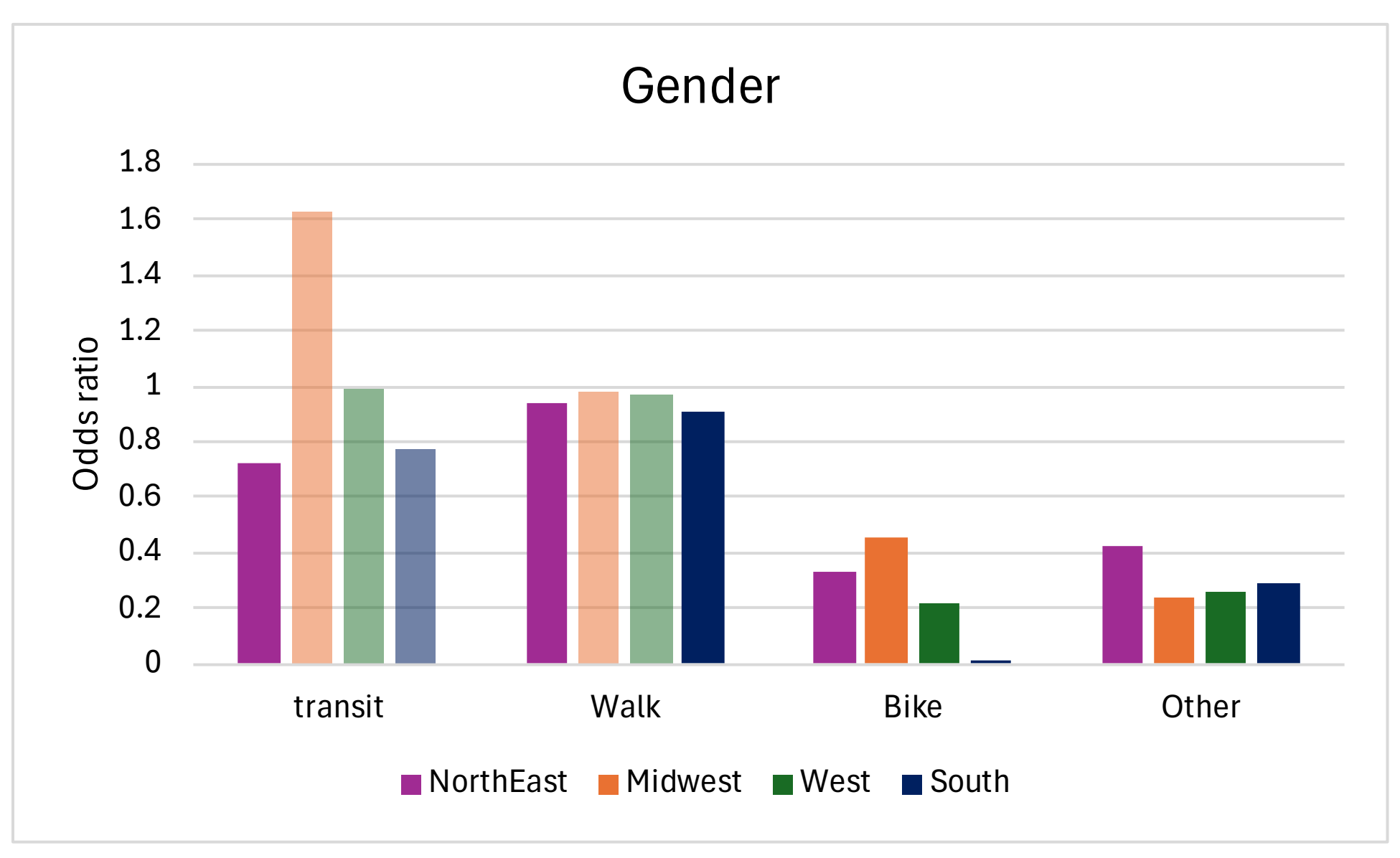

Education Level

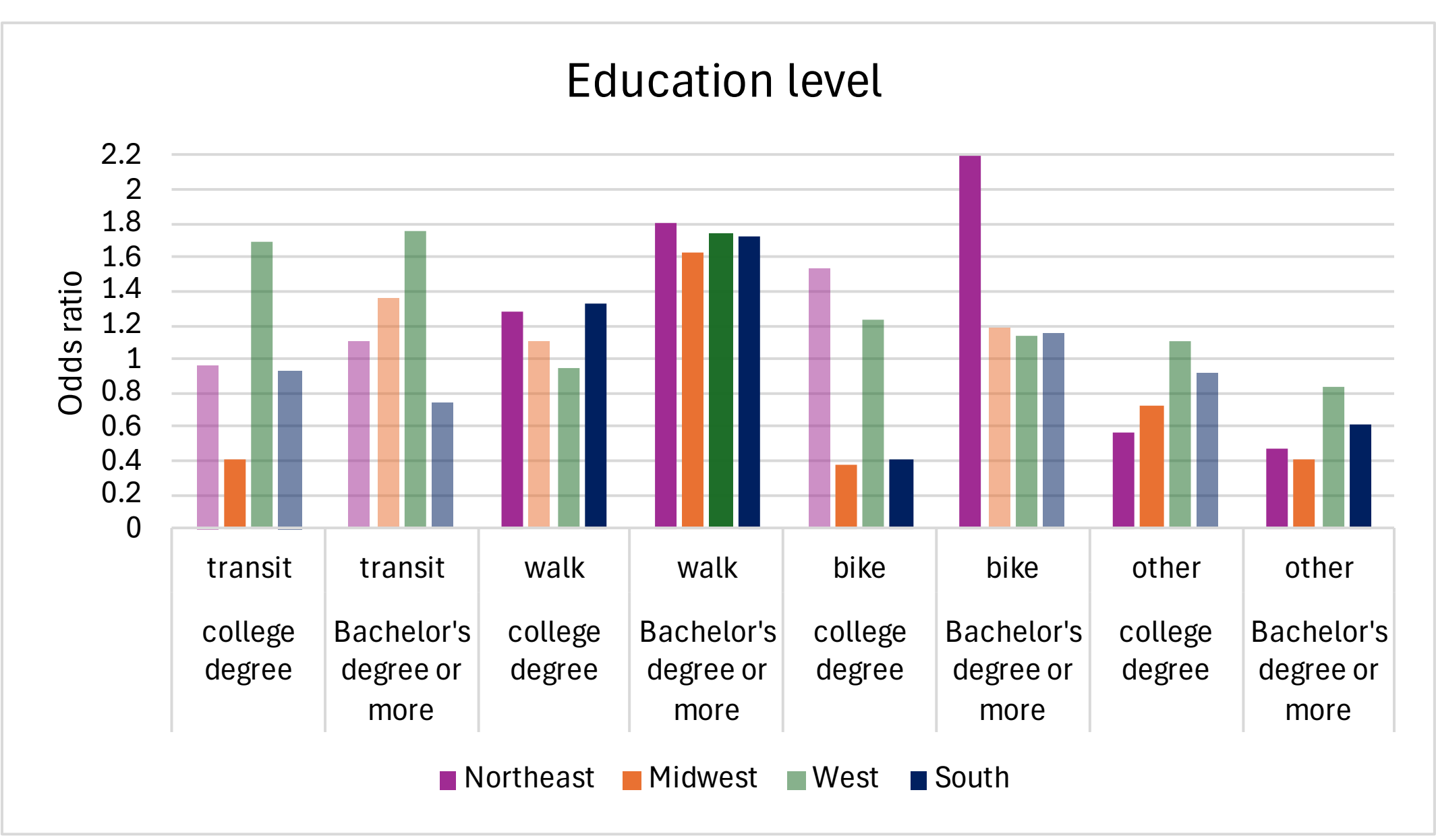

**Visualized Odds Ratios**

Age

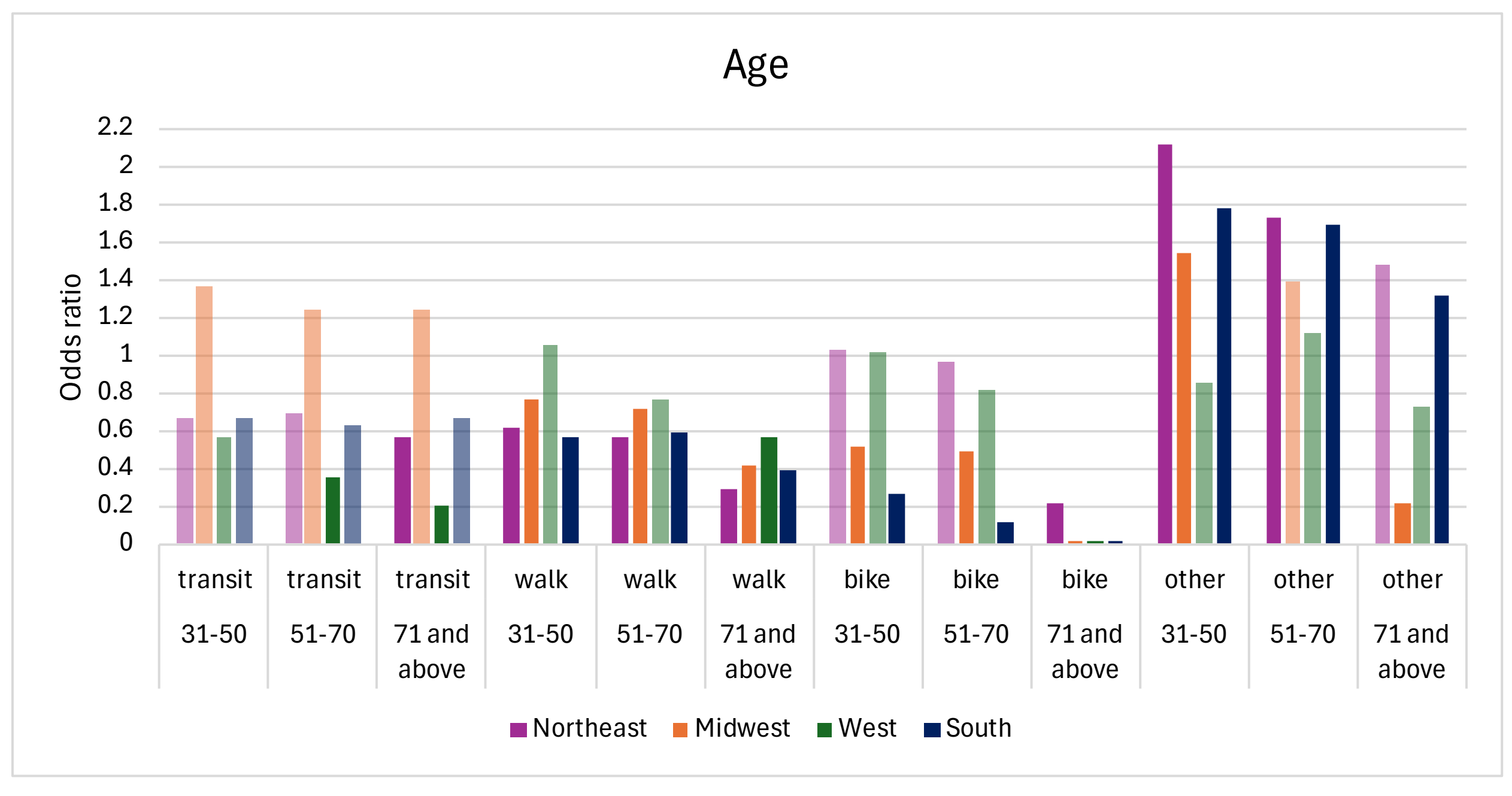

Race

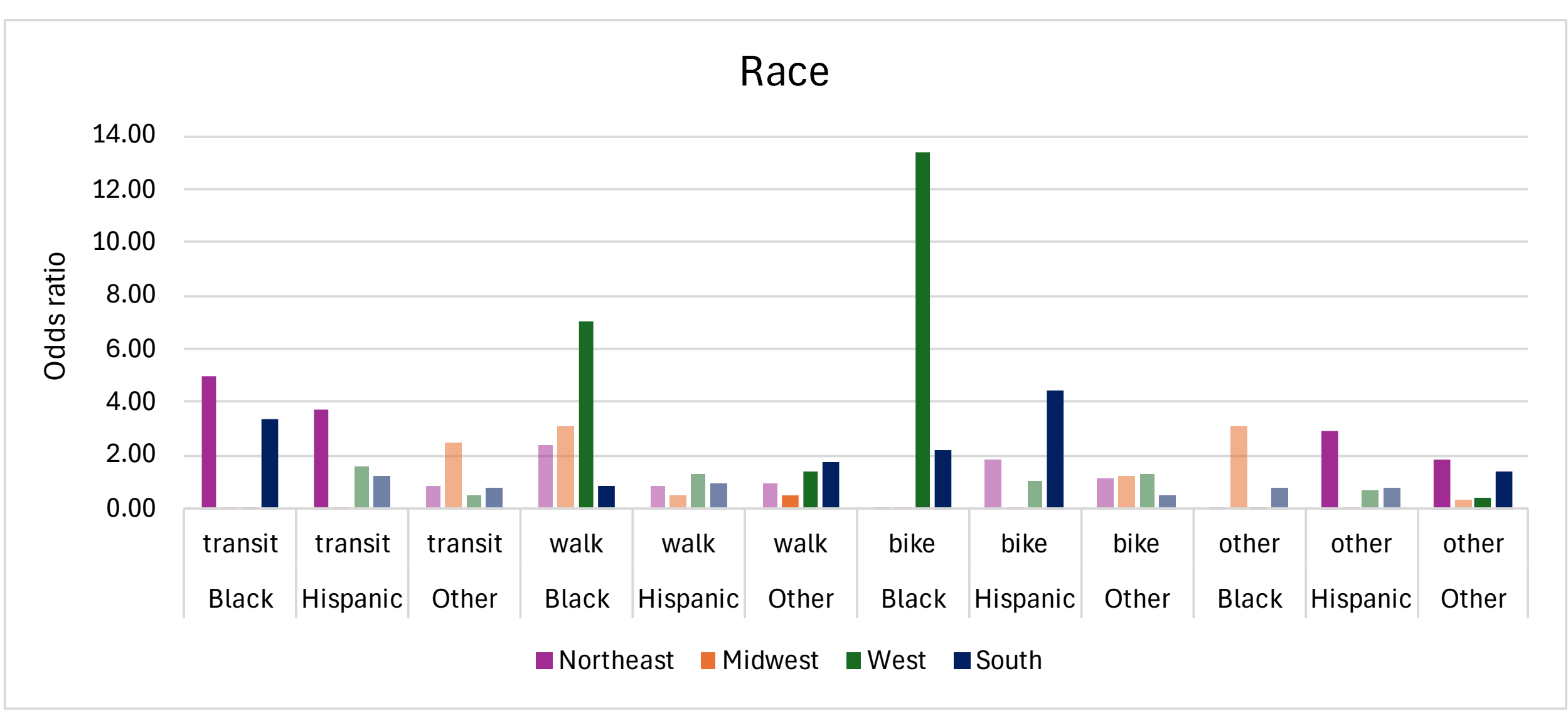

**Visualized Odds Ratios**

Household Size

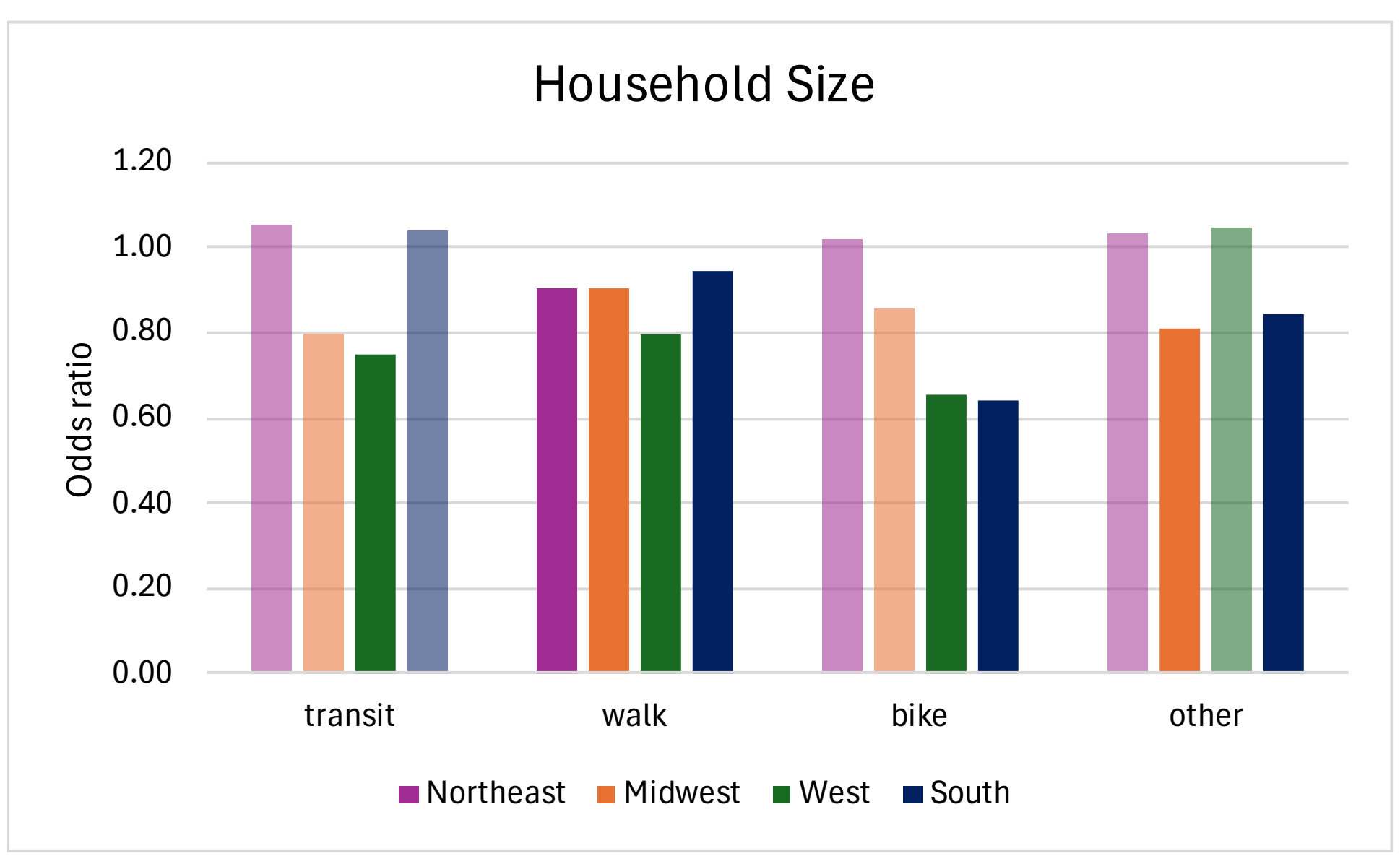

Household Vehicles

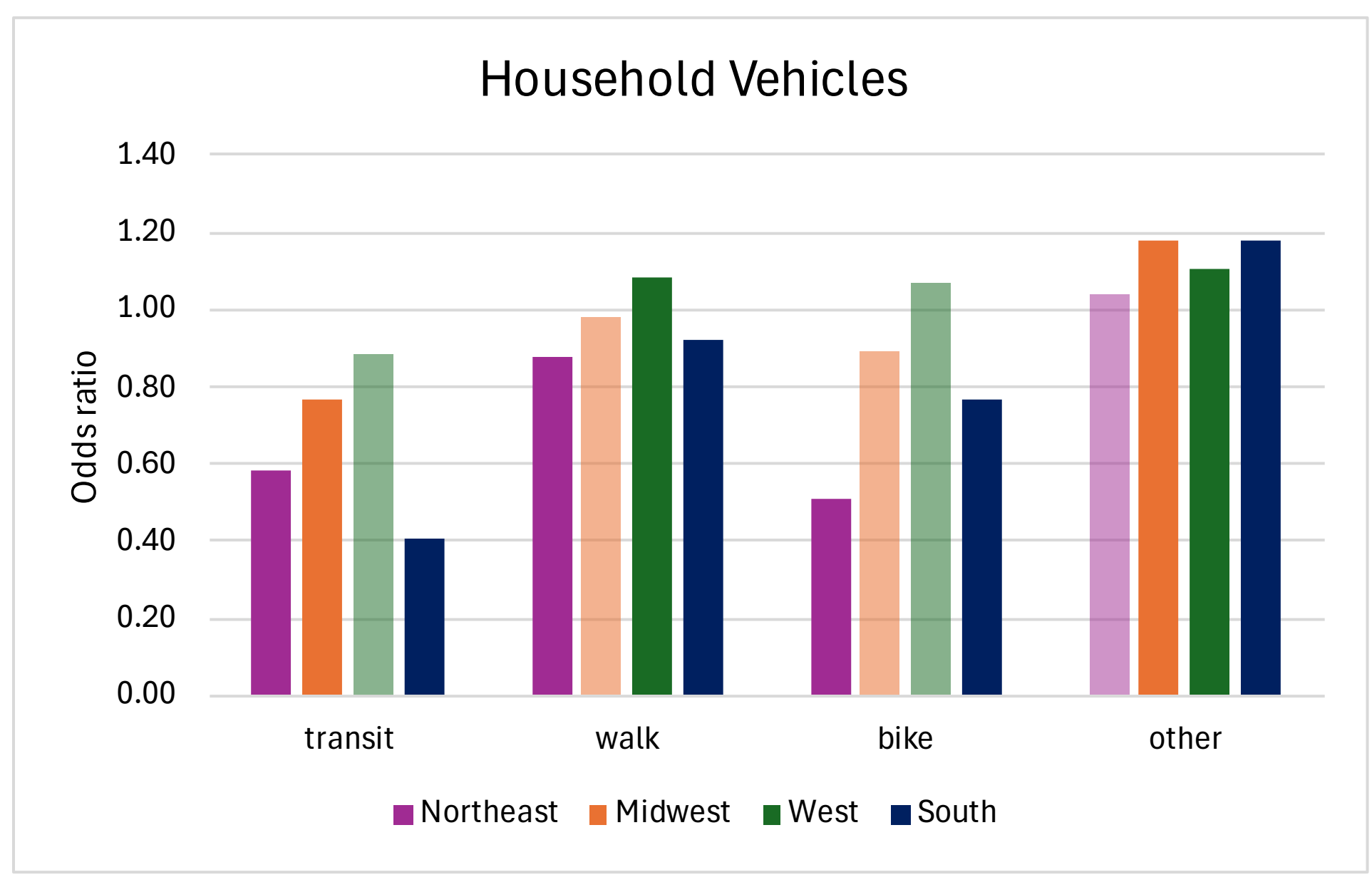

**Visualized Odds Ratios**

Trip Day

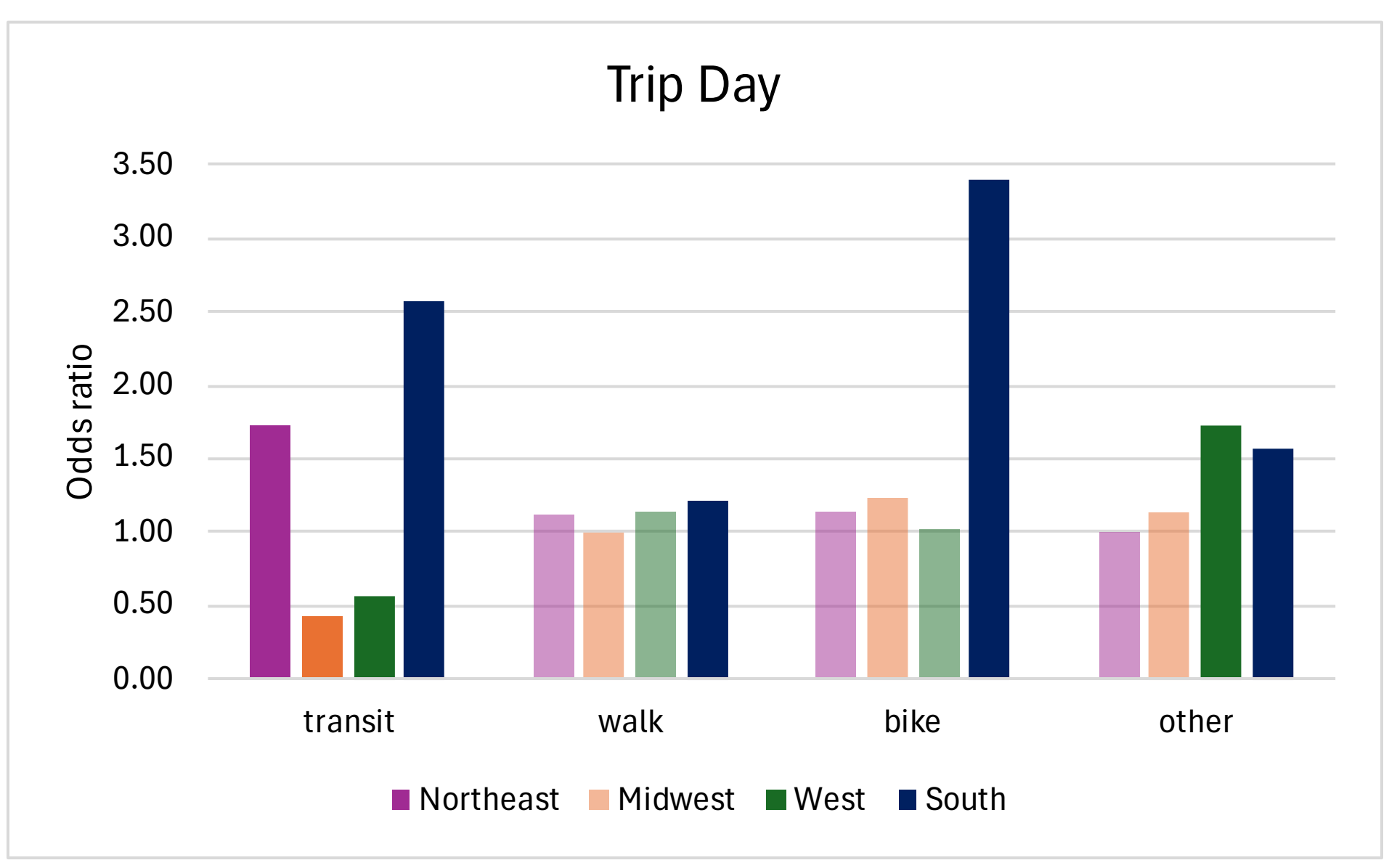

Trip Miles

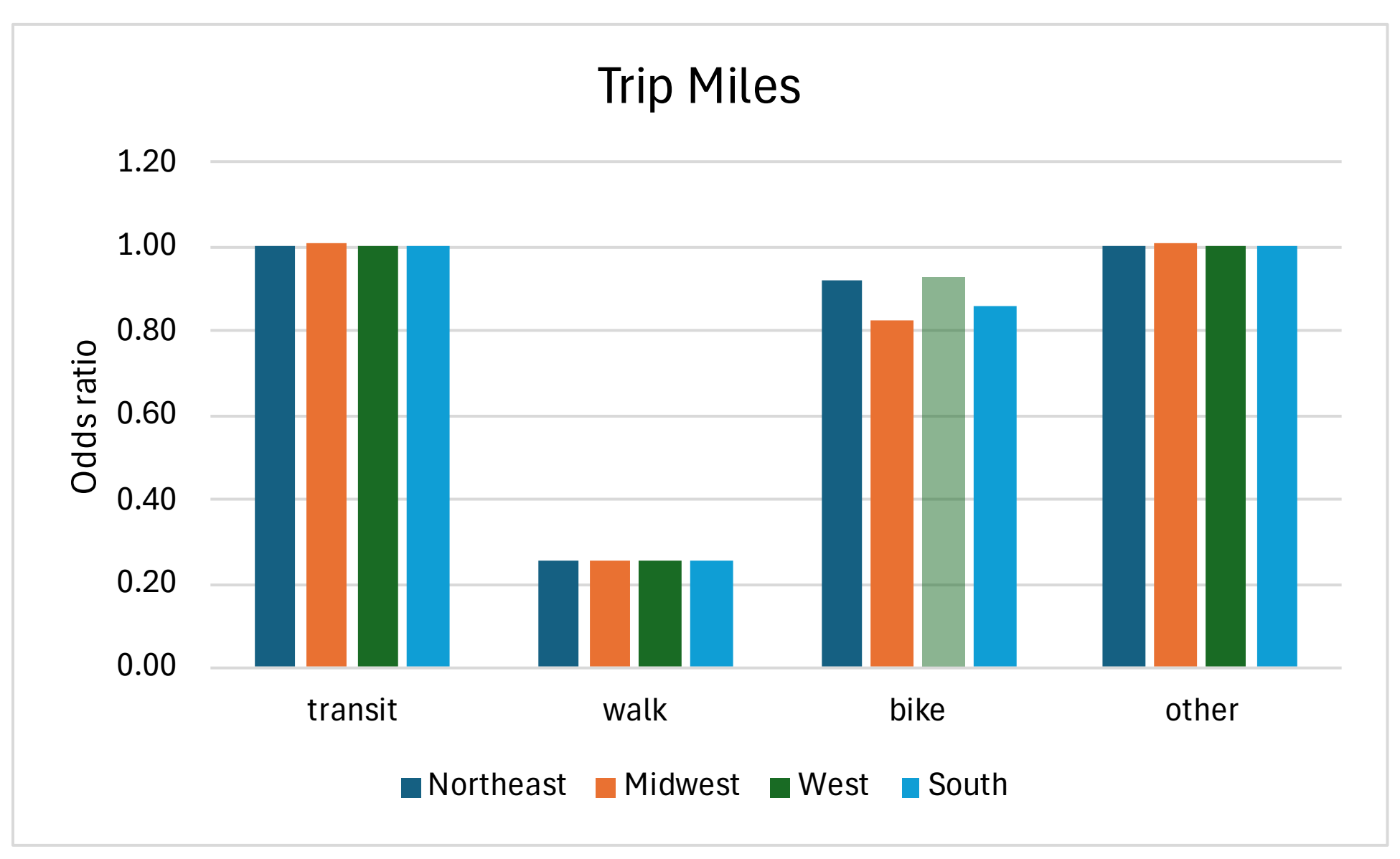